\documentclass[aps,prl,a4paper,twocolumn,superscriptaddress,showkeys,longbibliography]{revtex4-1}

\usepackage[colorlinks=true,bookmarks=false,citecolor=black,linkcolor=black,hyperfootnotes=true,urlcolor=black]{hyperref}

\usepackage{physics}
\usepackage{graphicx}
\usepackage{dcolumn}
\usepackage{bm}
\usepackage[dvipsnames]{xcolor}
\usepackage{wasysym}
\usepackage{lipsum}
\usepackage{enumitem}
\usepackage{booktabs} 
\usepackage{gensymb}
\usepackage{xcolor}
\usepackage{amssymb}
\usepackage{booktabs}
\usepackage{array}
\newcolumntype{L}{>{$}l<{$}}
\newcolumntype{C}{>{$}c<{$}}
\newcolumntype{R}{>{$}r<{$}}

\usepackage{mathtools}
\usepackage{siunitx}
\usepackage{soul}
\usepackage{float}

\makeatletter
\def\p@subsection{}
\makeatother

\begin{document}

\title{Absence of a BCS-BEC crossover in the cuprate superconductors}

\author{John Sous} \email{sous@stanford.edu}
\affiliation{Department of Physics, Stanford University, Stanford, California 94305, USA}

\author{Yu He} \email{yu.he@yale.edu}
\affiliation{Department of Applied Physics, Yale University, New Haven, Connecticut 06511, USA}

\author{Steven A. Kivelson} \email{kivelson@stanford.edu }
\affiliation{Department of Physics, Stanford University, Stanford, California 94305, USA}

\date{\today}

\begin{abstract} 
We examine key aspects of the theory of the Bardeen-Cooper-Schrieffer (BCS) to Bose-Einstein condensation (BEC) crossover, focusing on the temperature dependence of the chemical potential, $\mu$.  We identify an accurate method of determining the change of $\mu$ in the cuprate high temperature superconductors from angle-resolved-photoemission data (along the `nodal' direction), and show that $\mu$ varies by less than a few percent of the Fermi energy  over a range of temperatures from far below to several times above the superconducting transition temperature, $T_c$.  This shows, unambiguously, that not only are these materials always on the BCS side of the crossover (which is a phase transition in the $d$-wave case), but are nowhere near the point of the crossover (where the chemical potential approaches the band bottom).
\end{abstract}

\maketitle

\normalsize
\vspace{-2mm}
\section{Introduction}
\begin{figure*}
\centering
\includegraphics[width=1.9\columnwidth]{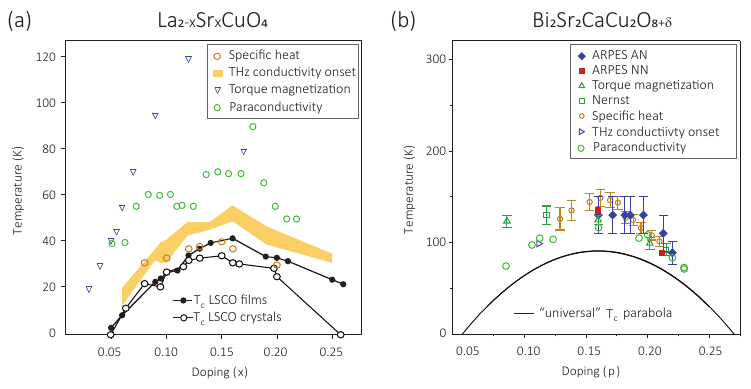}
\vspace{-2mm}\caption{{\bf Cuprates phase diagram.} Phase diagram of two representative cuprates, (a) La$_{2-x}$Sr$_x$CuO$_4$ (LSCO) and (b) Bi$_2$Sr$_2$CaCu$_2$O$_{8+\delta}$ (Bi-2212), as a function of doped hole concentration and temperature. Open  black markers indicate $T_c$ for bulk crystals, while the solid black line is the same for crystalline films from Ref.~\cite{bilbro2011temporal}.  The other symbols indicate crossover lines below which the existence of significant superconducting fluctuations are inferred  from various different experiments that are directly sensitive to Cooper pair formation.
Data, including error bars where applicable, are reproduced from Refs.~\cite{bilbro2011temporal,chen2019incoherent,he2021superconducting,tallon2011fluctuations,corson1999vanishing,usui2014doping,ando2004electronic,li2010diamagnetism,wang2005field,kondo2015point,martin}. The quantitative identification of any crossover   depends on the criterion used, and moreover distinct probes should have different sensitivity to superconducting correlations, so it is reasonable that the various lines do not coincide.}
\label{fig:phaseDiagram}
\end{figure*}

\vspace{-4mm}
The zero temperature ($T$) superfluid density, $n_s(0)$, of the cuprate high temperature superconductors is several orders of magnitude smaller than that of conventional superconductors~\cite{uemura1989universal,uemura1991basic,emery1995importance}. Indeed (when translated into energy units) it is comparable to the critical transition temperature ($T_c$)~\cite{emery1995importance}.  This has led to the probably inescapable inference that $T_c$, itself, is determined, at least to a significant degree, by the condensation scale (i.e. the phase ordering temperature, $T_\theta \propto n_s(0)$), rather than by the pairing scale ($\sim  \Delta_0/2$), in contrast to the case in the Bardeen-Cooper-Schrieffer (BCS) theory of conventional superconductors. There is also compelling evidence that some degree of clearly identifiable  superconducting fluctuations - colloquially referred to as `pairing without phase coherence' - persists for a substantial range (at least 20\% or so) above $T_c$~\cite{wang2005field,li2010diamagnetism,kondo2015point,chen2019incoherent,he2021superconducting,corson1999vanishing,bilbro2011temporal,hu2014optically,bergeal2008pairing,zhou2019electron,bovzovic2016dependence}.  This has been known for some time for the underdoped cuprates, but it has recently become increasingly clear that the same is true for many or all overdoped cuprates as well (Fig.~\ref{fig:phaseDiagram})~\cite{he2021superconducting,zhou2019electron,tromp2022non}.  Indeed, as a function of doping,  the onset temperature (however defined) of superconducting fluctuations more closely parallels $T_c$ than it does the conventionally defined pseudo-gap crossover. 

However, what is unclear is why this occurs, and what we should learn from this.  One proposal is that this should be taken as evidence that the system is approaching a strong pairing situation, referred to as the Bose-Einstein condensation (BEC) limit, in which the electrons form non-overlapping charge 2$e$ bosons at a scale far above $T_c$~\cite{TDLee,Randeria1,Randeria2,PhysRevLett.71.3202,AlexandrovMott,KLevin1,KLevin2,KLevin3,randeria2014crossover,harrison2022magic,KLevinReview,LevinNote}.  However, as we will discuss below, there are other conceptually distinct, yet equally well understood circumstances in which $T_c$ is determined by phase ordering and in which Cooper pairing persists above $T_c$.  The purpose of this work is to analyze the behavior that would be expected of a system either in the BEC limit or approaching the BCS to BEC crossover from the BCS side, and to present direct experimental evidence that this is {\em not} the case for the cuprates.

\vspace{-2mm}
\section{Framing the issue}\vspace{-2mm}

BCS theory is a weak coupling theory that is built on a starting point that is the electronic structure from band theory. The BEC limit invokes electronic bound states.  In the former case, the pairing is highly collective and the chemical potential, $\mu$, is only weakly affected by the advent of pairing.  In the latter, the chemical potential - by the definition of a bound state - must approach a value that lies below the band bottom as $T\to 0$.  These  differences do not refer to subtle low-energy phenomena but rather to entirely different regimes of microscopic physics on energy scales of the order of the Fermi energy, $E_F$, or larger~\cite{de2008fermions}. 

\begin{figure}[!b]\vspace{-4mm}
\centering
\includegraphics[width=\columnwidth]{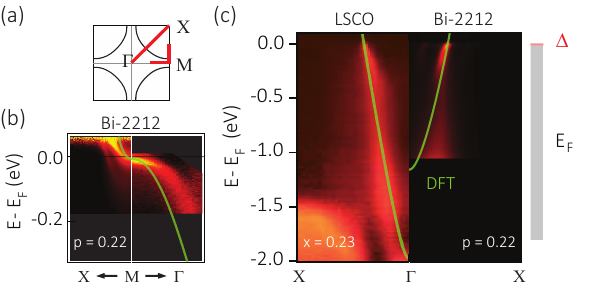}
\caption{{\bf Electronic structure of Bi-based cuprates along high symmetry directions.} (a) Schematic Fermi surface and momentum cut trajectory in the tetragonal Brillouin zone of a CuO$_2$ plane. (b) Low energy electronic structure near ($\pi$,0) in Bi-2212 ($\mathrm{p} = 0.22$, $T_c = 66$ K) in the normal state. Data are reproduced from Ref.~\cite{he2021superconducting}. (c) Electronic structure along $\Gamma-X$ direction in LSCO ($x = 0.23$, $T_c = 24$ K) and Bi-2212. Light green lines are density functional theory (DFT) calculated band structure. For Bi-2212 only the antibonding band is shown. Deviations from the first principles dispersion apparent at low energies represent mass renormalization due to additional interaction effects, see Supplementary Note 9.
Data are adapted from Refs.~\cite{he2021superconducting,he2018spectroscopic,lin2006raising,kramer2019band}.}
\label{fig:bandStructure}
\end{figure}

From this perspective, the fact that the Fermi surface and general features of the electron dispersion seen in ARPES experiments across the superconducting dome of the cuprates are more or less in agreement with expectations from band-structure calculations appears to be inconsistent with any large excursions toward the BEC limit.  (This is illustrated in Fig.~\ref{fig:bandStructure}.)  
Emergent features of the low energy physics, such as a normal state pseudo-gap that competes with superconductivity (apparent below some generally relatively ill-defined $T^\star$)~\cite{alloul198989,batlogg1994normal,loeser1996excitation,ding1996spectroscopic,renner1998pseudogap,daou2009linear,hashimoto2014energy,chen2019incoherent} and various low energy kinks in the dispersion relations are certainly interesting and important, but occur on energy scales small compared to $E_F$~\cite{lanzara2001evidence,zhou2003universal,vishik2010doping,anzai2017new}.  The fact that the application of magnetic fields large enough to quench superconductivity produces quantum oscillations~\cite{QOReview} is further evidence that pairing is a collective property of the superconducting state rather than a microscopic feature associated with bound-state formation~\cite{BoseQO}.

\footnotetext[3]{For a recent review. see Ref.~\cite{KeimerReview}.}

One important feature of the superconducting state in the cuprates is that it has $d$-wave symmetry and gapless, nodal quasi-particle excitations~\cite{Note3}.  There can be no nodal quasi-particles in the BEC limit. (See Ref.~\cite{duncan2000thermodynamic}.) Thus, for this $d$-wave case, the BCS to BEC crossover~\cite{randeria2014crossover,KLevinReview} would constitute a (Lifshitz) phase transition from a nodal to a nodeless state~\cite{duncan2000thermodynamic,botelho2005lifshitz}. The existence of well defined nodes is, of itself, proof that the cuprates are on the BCS side of the transition.  This leaves  only the question of how far on the BCS side they are from the point at which  a BCS to BEC transition would have occurred~\footnote{There are photoemission experiments reporting gapped nodes in highly underdoped cuprates below $\mathrm{p} = 0.08$ - $0.10$~\cite{peng2013disappearance,vishik2012phase,razzoli2013evolution,he2018spectroscopic}. However, these gaps preserve the nodal Fermi momenta, lack coherent quasi-particles, and are all two orders of magnitude smaller compared to the Fermi energy.}.  In the language of effective field theories, the question we consider is not one concerning the correct infra-red description (i.e. phases of matter) but rather concerns the ultra-violet (high energy `microscopic') description consistent with experimental data.

\vspace{-2mm}
\section{Experimental perspective}\vspace{-2mm}

\begin{figure}[!b]
\centering
\includegraphics[width=\columnwidth]{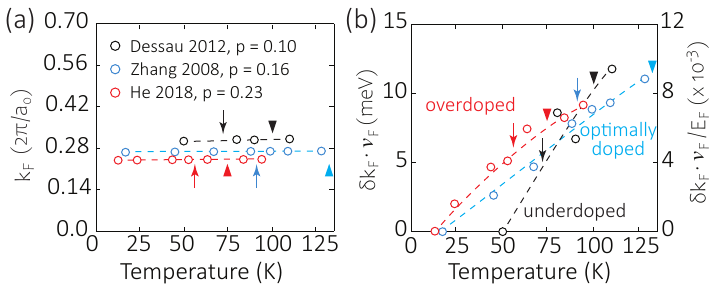}
\caption{{{\bf Temperature dependence of $k_F$ along the nodal direction.} (a) Bi-2212 nodal momenta normalized to $\frac{2\pi}{a_0}$, where $a_0$ = 3.8~\AA~is the in-plane Cu-Cu separation under tetragonal approximation. Data are reproduced from Refs.~\cite{reber2012origin,zhang2008identification,he2018spectroscopic}. (Note that if these materials were in the crossover regime,  the chemical potential would drop to the band bottom, and consequently $k_F$ would  shift to zero.) (b) Chemical potential shifts $\delta \mu = v_F \delta k_F$, evaluated by multiplying the $k_F$ shifts by the Fermi velocity~\cite{vishik2010doping} in absolute energy units and in units of $E_F = 1.25$~eV. In both panels, arrows indicate the corresponding values of $T_c$. Solid triangles (which correspond to the solid symbols in Fig.~\ref{fig:phaseDiagram}(b)) represent the superconducting gap opening temperature as determined from ARPES~\cite{chen2019incoherent,he2021superconducting}. Dashed lines are guides to the eye. Typical error bars (not shown) are  $\leq$ 4 meV (momentum resolution multiplied by Fermi velocity).
}}
\label{fig:experiment}
\end{figure}

We will focus our attention on the behavior of the chemical potential, $\mu$, as this is a fundamental thermodynamic quantity that  exhibits qualitatively different behavior in the two limits. 
Since by definition, in the BEC limit the chemical potential is below the band bottom, on approach to the BEC limit from the BCS side one should see that the chemical potential is significantly depressed from its band theory value toward the bottom of the band. Moreover, it should show strong $T$ dependencies for temperatures of order $T_c$. 

A number of fortuitous features of the electronic structure of the cuprates make it possible to stringently bound the evolution of the chemical potential from the electron dispersion measured in ARPES along the `nodal direction' in the Brillouin zone.  Specifically, it is possible to determine the value of the Fermi momentum, $k_F=|\vec{k_F}|$, as a function of temperature with a high degree of precision.  In the superconducting state, since the gap vanishes along this direction, it is possible to measure the quasi-particle dispersion to where it crosses the chemical potential.  Moreover, since the pseudo-gap - where it exists - also has a $d$-wave structure~\cite{Note3}, it also vanishes along this direction. Working further to our advantage is the fact that this is the trajectory through the Brillouin zone along which the spectral peaks seen in ARPES are the sharpest.  Indeed, in bilayer cuprates (such as Bi-2212), the quantum chemistry results in a vanishing  bilayer splitting at the node, so there is no need to worry about this complication either.

While knowledge of the temperature evolution of nodal $k_F$ does not permit an absolute measure of $\mu$, it does allow a direct measure of changes in the chemical potential relative to a reference value,
\begin{align}
    \delta \mu(T) = v_F \delta k_F(T),
\end{align}
where $v_F$ is the nodal velocity (which, conveniently, appears to be minimally dependent on $T$ and on doped hole concentration, $x$, on the 10 - 100 meV energy scale~\cite{zhou2003universal,vishik2010doping}). It should be noted that a cascade of nodal dispersion kinks can affect the temperature dependent shift of the nodal $k_F$ up to 1\% of the reciprocal lattice unit~\cite{lanzara2001evidence,zhou2003universal,vishik2010doping,anzai2017new}, which is comparable to experimentally observed values, but at least 2 orders of magnitude smaller than what is expected in the crossover regime.

In Fig.~\ref{fig:experiment}  we  show $k_F(T)$ from ARPES data in Bi-2212 for several different values of $p$. The measured changes in $k_F$ are sufficiently small that, within the uncertainties of interpretation, they are consistent with a temperature independent chemical potential. (Specifically, minute but difficult to quantify shifts of $k_F$ are expected to arise from temperature dependent low-energy band renormalizations due to electron-electron or electron-phonon interactions, even in the absence of any $T$ dependence of $\mu$.) This is our primary finding. Indeed below, and in Fig.~\ref{fig:theory}, we show that the very small changes in the chemical potential expected on the basis of BCS theory are order-of-magnitude consistent with these findings, assuming reasonable values of the $T=0$ gap, $\Delta_0 \approx 40$~meV.  By contrast, even on the BCS side of the transition, in the regime proximate to a BEC limit a large shift in the chemical potential toward the band bottom would be expected.

There are a few aspects of the result that merit closer inspection.  As can be seen in Fig.~\ref{fig:bandStructure}c, $E_F$ (defined to be the position of the band-bottom at the $\Gamma$ point relative to the chemical potential) is between 1 - 2 eV, which is  large enough compared to $\Delta_0$ that it would seem obvious that the system is deep in the BCS limit.   However, the band is relatively shallow near the van-Hove point. The energy at the $M$  point, $(0,\pi)$, is no more than 0.1~eV below $\mu$ (Fig.~\ref{fig:bandStructure}b)~\cite{yoshida2006systematic,he2018rapid}.  This  is only a few times $\Delta_0$, so that if we focused exclusively on this near `antinodal' region of the Brillouin zone, we might have anticipated more in the way of a shift in the chemical potential.  The idea that the anti-nodal `heavy electrons' can be viewed as somehow distinct from the near-nodal `light electrons', however, runs up against the experimental fact that the chemical potential does not show any of the $T$ dependence across $T_c$ or $T^*$ that should be a corollary of such a two-patch theory (see next section). Indeed, the bound we have obtained on the chemical potential shifts are so stringent so that  $|\delta \mu| < \Delta_0$. In addition, near the BCS-BEC crossover, the backbending momenta of the Bogoliubov quasiparticle dispersion in the superconducting state should shift towards zero, which is not observed in the cuprates at any hole doping~\cite{hashimoto2010particle,he2021superconducting,chen2022unconventional}.

\vspace{-2mm}
\section{Theory of the BCS to BEC crossover}\vspace{-2mm}
As in the experimental discussion, we  focus our theoretical analysis on the thermal evolution of the chemical potential $\mu$.  Specifically, we illustrate the fact that variations of $\mu$ are small in the BCS limit, increase upon approach to the BCS to BEC crossover, and are large whenever the BEC perspective is relevant~\footnote{One should note that there exists fine-tuned circumstances - both in terms of electron density (i.e. `charge-neutrality') and  band-structure considerations - under which an exact particle-hole symmetry pins the chemical potential to a specific value, independent of $T$ or whether one is in the normal or superconducting state.  Obviously in this case, the chemical potential cannot be used as a metric of the BCS to BEC crossover.  Such a symmetry is manifestly absent in the cuprates.}.
\vspace{-2mm}

\subsection{The BCS analysis}\vspace{-2mm}
Given that the existence of nodal quasi-particles places the cuprates on the BCS side of the transition, it is reasonable to consider signatures of the approach to the BEC limit in the context of BCS theory. It is an often neglected feature of BCS theory that, in addition to the familiar gap equation, there is a second self-consistency equation that determines the chemical potential as the implicit solution to
\begin{align}
    n= 2\int \frac {d\vec k}{(2\pi)^d} \left[|u_{\vec k}|^2(1-f_{\vec k}) +|v_{\vec k}|^2 f_{\vec k} \right]
\end{align}
where $n$ is the electron density, $f_{\vec k}=[e^{\beta E(\vec k)} +1]^{-1}$ is the Fermi function, $E(\vec k) = \sqrt{[\epsilon(\vec k) - \mu]^2 +|\Delta(\vec k)|^2}$ and $\epsilon(\vec k)$ are the quasi-particle energies in the superconducting and normal state respectively, $\Delta(\vec k)$ is the gap function, with the coherence factors $|u_{\vec k}|^2=[E(\vec k) -\epsilon(\vec k) +\mu]/2E(\vec k) $ and $|v_{\vec k}|^2=[E(\vec k) +\epsilon(\vec k) -\mu]/2E(\vec k) $, and $\beta = 1/T$.  If there are multiple bands, then this expression needs to be generalized to include a sum over bands.  Naturally, $\Delta$ depends implicitly on $T$ and on the nature of the interactions through the usual self-consistency relation.

To illustrate why this equation is safely neglected in most cases, consider the illustrative example of free electrons $(\epsilon(\vec k) = \hbar^2 k^2/2m$) in two dimensions ($d=2$) with a $\vec k$ independent ($s$-wave) gap function.  Because the density of states is constant, the integrals above can be performed readily, with the result that $\mu(T,\Delta)$ is obtained as the implicit solution to
\begin{align}
    E_F =&  \frac 1 2 \left[E_\mu+\mu \right]\nonumber+T\ln\left[1+e^{-\beta E_\mu}\right] 
\end{align}
where the Fermi energy  $E_F=\pi \hbar^2 n/m$ such that $\mu(T,\Delta=0)\to E_F$ as $T\to 0$ and $E_\mu = \sqrt{\mu^2 + \Delta^2}$.  The second term in this equation is typically exponentially small, $\sim e^{-\beta E_F}$, and hence 
\begin{align}
    \mu(T,\Delta) = E_F\left[1 - |\Delta(T)/2E_F|^2\right] +{\cal O}(T e^{-\beta E_F}).
    \label{Fermi}
\end{align}
If we define the BCS to BEC crossover as the point at which $\mu(0,\Delta)=0$,  this occurs when $\Delta_0 \equiv \Delta(0) = 2E_F$ - the shift of the chemical potential relative to its normal state, $[\mu(T>T_c) - \mu(0)]/\mu(T>T_c)\approx [E_F - \mu(0)]/E_F$,  thus is directly a measure of how closely we have approached this crossover.

In more general circumstances, band-structure effects result in an energy dependent density of states.  In the small $\Delta_0$ (BCS) limit, this leads to a (logarithmically) larger shift in the chemical potential  $\delta \mu = \mu(0,\Delta_0)-\mu(0,0)$ of the form
\begin{align}
    \delta \mu \sim -\frac{1}{2} \frac{\rho^\prime(\mu(0,0))}{\rho(\mu(0,0))} |\Delta_0|^2 \ln \left(\frac{2W}{\Delta_0}\right),
    \label{vHshift}
\end{align}
where $\rho^\prime(\mu)$ is the derivative of $\rho$ with respect to $\mu$ and $W$ is the electronic bandwidth. On the other hand, if we continue to follow the evolution of $\mu$ according to the BCS equations for a single band to the large $\Delta_0$ limit (where, of course, BCS theory is not in any way justified) the result is asymptotically independent of the band-dispersion:
\begin{align}
     \mu = \left[ \frac{(n-1)}  {  \sqrt{n(2-n)}}\right ]   |\Delta_0|\left[ 1 +{\cal{O}}\left(\frac W {|\Delta_0|}\right)\right].
    \label{mubottom}
\end{align}
For a cuprate-like band-structure, these two asymptotic forms typically give rise to a non-monotonic dependence of $\mu$ on $\Delta_0$.  The presence of a van-Hove point below the Fermi energy implies that $\rho^\prime(\mu)$ is negative, meaning that for small $\Delta_0$, the superconductivity induced changes in the chemical potential are expected to be positive.  On the other hand, for a hole-doped cuprate (with $(n-1)<0$), the chemical potential must drop toward the band bottom for large enough $\Delta_0$.

\begin{figure}[!t]
\centering
\includegraphics[width=\columnwidth]{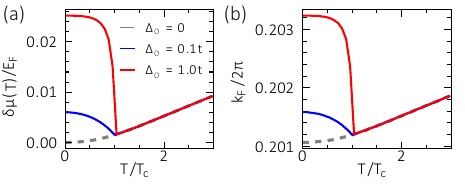}
\caption{{\bf Thermal evolution of the chemical potential in the BCS limit.} Thermal evolution of (a) $\delta\mu(T) = \mu(T)-E_F$ and (b) $k_F$ computed for the model of a $d$-wave BCS superconductor discussed in the text.   The grey dashed line is for the normal state $\Delta_0 = 0$, and the blue and red solid lines are, respectively, for $\Delta_0 = 0.1t$ and $\Delta_0 = t$. We have taken  a density of doped holes $x=0.2$ ($n=0.8$) and $T_c = 0.025t$. Temperatures are shown in units of $T_c$ and energies in units of $E_F = 1.7t$.}
\label{fig:theory}
\end{figure}

To make closer contact with experimental reality, we have numerically carried through the BCS analysis 
for a two-dimensional (2D) model that incorporates significant features of the electronic structure of the cuprates.  Here, we take  $\epsilon(\vec k)=4(t+t')-2t (\cos(k_x) +\cos(k_y))  -4t'\cos(k_x) \cos(k_y)$ (the zero of energy has been chosen to coincide with the band bottom, i.e. such that $\epsilon(\vec 0)=0$ and we have set the lattice constant $a=1$) with $t'=-0.3t$ (obtained from a fit to low binding energy ARPES data~\cite{he2021superconducting,chen2022unconventional,Notetbfits}) and $\Delta(\vec k)= \Delta(T)[{\cos(k_x)-\cos(k_y)}]/2$, where we further assume that $\Delta(T) = \Delta_0\sqrt{1-(T/T_c)^2}$ for all $T<T_c$. We perform the calculation for hole concentration $x=0.2$ ($n=1-x=0.8$) and we use $T_c = 0.025t$.  The results are shown in Fig.~\ref{fig:theory} for two values of $\Delta_0$: 1) $\Delta_0=0.1t$, which is a reasonable value for the cuprates; 2) $\Delta_0 = t$, which is far larger than is plausible, included for illustrative purposes. Not only is the thermal evolution of $\mu(T)$ very weak relative to $E_F$, but also it appears that it is very small relative to $\Delta(T)$. Note that this analysis is \textit{not} meant to quantitatively explain the experimental results, but rather to provide an estimate of the expected magnitude shift in $\mu$ and $k_F$ across $T_c$ in the BCS limit.

\vspace{-2mm} 

\begin{figure}[!t]
\centering
\includegraphics[width=\columnwidth]{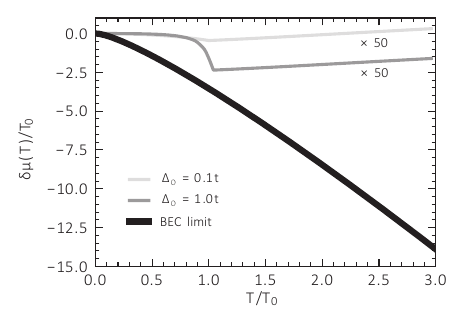}
\caption{{\bf Thermal evolution of the chemical potential in the BEC limit.} Thermal evolution of $\delta\mu(T) = \mu(T) - \mu(0)$ in the BEC limit (black thick line) contrasted with its behavior in the BCS limit for $\Delta_0 = 0.1t$ (light grey thin line) and $\Delta_0= t$ (dark grey thin line). As in Fig.~\ref{fig:theory}, in the BCS calculation we have taken $T_c = 0.025t$ and a density of doped holes $x=0.2$ ($n=0.8$),  which yields $E_F = 1.7t$. In comparing the BEC and BCS results we identified energy scales according to $T_c\equiv T_0$,  $E_F\equiv 2T_0$, and  scaled up the BCS results by a factor of $50$ in order to make the $T$ dependence visible.}
\label{fig:theory2}
\end{figure}

\subsection{The BEC limit}\vspace{-2mm}
To develop intuition concerning the thermal evolution of $\mu$ in the BEC limit, we can carry out the  same analysis for the case of a 2D non-interacting Bose gas. While this problem has no actual phase transition, in the presence of weak repulsive interactions, the superfluid transition occurs at a number of order 1 (which depends on the log-log of the interaction strength~\cite{fisher1988dilute}) times the characteristic energy $T_0 \equiv 2\pi \hbar^2 n_B/m_B$. (If we identify the areal density  of the bosons, $n_B$  with 1/2 the density of Fermions, $n$, and  their mass $m_B$ as twice the electron effective mass, $m$, then $T_0=E_F/2$, where $E_F$ is what would have been the Fermi energy in the absence of pair-binding.) Again, the fact that the density of states is a constant permits us to derive an analytic expression 
\begin{align}
    \mu = T \ln[1-e^{-T_0/T}].
    \label{bose}
\end{align}
This result is shown in Fig.~\ref{fig:theory2}. From Eq.~\ref{bose}, it is easy to see that the chemical potential shifts by approximately a factor of 4 as the temperature changes from $T=T_0$ (roughly $T_c$) to $T=2T_0$; a result that  is qualitatively unchanged by weak interactions.\vspace{-2mm}

\subsection{Mixture of heavy bosons and light fermions}\vspace{-2mm}
Motivated by the proposal of heavy antinodal electron pairs mixing with light nodal quasiparticles in cuprates, it is interesting to consider a two component system in which a 2D Bose gas is in equilibrium with a BCS superconductor~\cite{TDLee}.  One could imagine this arising in a two-band system, in which one band is in a BCS and the other in a BEC limit.  Now the chemical potential must simultaneously satisfy Eqs.~\eqref{Fermi} and \eqref{bose} - which in turn means that the fraction of particles that are bosonic must be determined self-consistently according to
\begin{align}
    n_{tot} = n + 2n_B = (1/\pi \hbar^2)\left[ m E_F +m_B T_0\right]
\end{align}
where $n_{tot}$ is the total electron density, and the factor $2$ encodes the assumptions that two electrons can combine to form one boson.  The result is a generally complicated thermal evolution of $\mu$.  However, in the limit that $m_B \gg m$ (i.e. where the bosonic density of states is large compared to the fermionic density of states), the result simplifies; here, the density of fermions does not change significantly over the relevant range of $\mu$, so the $T$ dependence of $\mu$ reduces to the same expression as for the pure bosonic problem, Eq.~\eqref{bose}, with an approximately constant value of $n_B$.

\section{Quantifying Cooper pair overlap}

One line of analysis that is sometimes invoked in support of proximity to a BEC limit is based on an estimate of the number of Cooper pairs in a Cooper pair area~\cite{KLevinReview,Leggett2006NatPhys}. This is estimated as $N \equiv n \pi |\xi_0|^2$, where $n$ is the density of conduction electrons per unit area and $\pi|\xi_0|^2$ represents the area associated with a given pair ($\xi_0$ is the correlation length).   It is then proposed that $N$ is a reasonable metric, such that $N \gg 1$ in the BCS limit and $N\lesssim 1$ in the BEC limit. 

However, neither $n$ nor $\xi_0$ is well defined.  For instance, in the cuprate context, there is an order of magnitude uncertainty concerning what value of $n$ is appropriate - whether it is proportional to $x$, the density of `doped holes' relative to the undoped insulator, or $(1+x)$, the area enclosed by the Fermi surface~\cite{x_and_EF}.  $|\xi_0|^2$ is even more uncertain, given that this is a nodal superconductor. The Fermi surface average of $|\xi_0|^2$ is infinite due to its divergence in the nodal direction.  Taking this at face value it suggests (not without reason) that a nodal SC can never approach the BEC limit.  On the other hand, it is the Fermi surface average of $\xi^{-2}$ that enters the mean-field estimate of $H_{c2}$, and this is dominated by the portions of the Fermi surface where the gap is maximal and/or the Fermi velocity is minimal.   If one makes an estimate of $N$ taking  the shortest possible estimate of $|\xi_0|^2$ (obtained from the largest experimentally inferred values of the mean-field $H_{c2}$) and the smallest possible value for $n \sim x$, the result suggests $N \sim 1-10$ for optimally doped cuprates - small enough that it might justify conjectures of a nearby BCS to BEC crossover.  However, because of the uncertainties  that lead to this estimate, we consider this analysis far less reliable than the analysis based on measurements of $\mu$.

\section{Further issues}

To complete our analysis of the physics of pairing in the cuprates, it is important to ask whether there are any alternatives to the BEC perspective that can account for the experimental observations of an intimate relation between  $T_c$ and  $T_\theta$ and a corresponding   persistence of pairing without phase coherence in an usually large range of temperatures  above $T_c$.  Three theoretically  understood  examples of systems that exhibit these properties are:
\begin{itemize}[leftmargin=*,noitemsep,topsep=0pt,parsep=0pt]
    \item A granular superconductor or Josephson junction array where $T_\theta$ (and hence $T_c$) is determined by the magnitude of the Josephson coupling between superconducting grains, while the pairing within a grain can be well described in the context of BCS theory~\cite{DynesGranular,pelc2019universal}. 
    \item A quasi-1D superconductor, where the pairing (gap formation) can occur in a BCS-like manner on a single superconducting wire, while $T_c$ is small in proportion to a positive power of the coupling between wires~\cite{1Dwires}.
    \item A lightly doped spin-liquid 
    of an appropriate variety, where the pairing scale is inherited from the spin correlations of the undoped insulator, while the superfluid density grows linear with doping, $x$~\cite{RKS,ChenKivelson}.
\end{itemize}

\noindent Which, if any of these possibilities is essential in the cuprates is still open to debate.  There is surely considerable evidence of significant  inhomogeneity in the electronic structure revealed by local probes~\cite{crocker2011nmr,jurkutat2019tc,pan2001microscopic,howald2001inherent,parker2010nanoscale}, so much so that there are suggestions that the cuprates should be viewed as electronic glasses~\cite{CuprateGlass1,CuprateGlass2,CuprateGlass3,CuprateGlass4,CuprateGlass5,tromp2022non,pelc2021unconventional}. In this light, it certainly is worth considering whether the materials might in some ways behave like granular superconductors.  While there is no direct evidence of either quasi-1D electronic structure, or of any spin-liquid phases - doped or otherwise, it is not obvious  (in the sense of adiabatic continuity) that these examples are totally irrelevant.  At the very least, the existence in the cuprate phase diagram of a variety of  `intertwined orders'~\cite{Note3}, especially charge-density-wave order, likely plays a role in reducing the fraction of the electrons that contribute to $n_s$. 
 
In concluding, we address two points of perspective concerning the present results that could easily be misinterpreted:
\begin{itemize}[leftmargin=*,noitemsep,topsep=0pt,parsep=0pt]
    \item The fact that the lack of substantial chemical potential shift with temperature is consistent with BCS theory does not prove that BCS theory is adequate to treat the emergent low energy properties of the cuprates.  It resolves the high energy microscopic issue of what are the constituent degrees of freedom one should include in a theoretical treatment - they are roughly the quasi-particles of a Fermi liquid and not preformed Cooper pairs.  However, the unusually large degree of superconducting fluctuations and the many other strange low energy behaviors of these materials certainly require more elaborate theoretical approaches than the simple BCS mean-field theory that works so well in conventional superconductors.
    \item  The idea that a BCS to BEC crossover may be at play has been mooted~\cite{DungHaiLee,BCSBECFeTe,lu2017zero,ma2021strongly,Iwasa,TBG,Shi_2022} in the context of a variety of other unusual superconductors, including  the Fe-based superconductors and more recently twisted bilayer graphene.  Obviously, the evidence that this crossover is not relevant in the cuprates does not prove that it is not significant in other materials.  Conversely, we propose that clear evidence can be obtained one way or the other from careful measurements of the evolution of $\mu$ as a function of $T$ and other properties that affect the superconducting state.
\end{itemize}

\section{acknowledgements}
We acknowledge useful discussions with I.~Bozovic,  V.~Calvera, L.~Glazman,  A.~Millis, D.~Natelson, C.~Sa~de~Melo,  and especially with T.~Devereaux, M.~Greven, M.~Hashimoto, D.-H.~Lee, K.~Levin, B.~Ramshaw, Z.-X.~Shen and J.~Tranquada. J.~S. acknowledges support from the Gordon and Betty Moore Foundation’s EPiQS Initiative through Grant GBMF8686 at Stanford University. Y.~H. acknowledges support from National Science Foundation under Grant DMR-2132343 at Yale University. S.~A.~K. acknowledges support from National Science Foundation under Grant DMR-2000987 at Stanford University.

\section{Data Availability} Experimental data are all retrieved from the published research articles referenced in the captions. The simulation data and code will be made available upon reasonable request.

\section{Author Contributions} All authors contributed to the inception, execution, and writing of this work.

\section{Competing Interest} The authors declare no Competing Financial or Non-Financial Interests.


\begin{thebibliography}{91}
\bibitem [{Lev()}]{LevinNote}%
  \BibitemOpen
  \href@noop {} {\ }\bibinfo {note} {The distinction between the BEC and BCS
  limits is theoretically unambiguous. However, there is no unique definition
  of the ``BCS to BEC crossover regime.'' Here, we have adopted a thermodynamic
  definition in terms of the change in the chemical potential. Others, notably
  in Ref.~\cite{KLevinReview}, have proposed a broader definition to include a
  range of conditions in which there is a significant fluctuational regime
  above $T_c$.}\BibitemShut {Stop}%
\bibitem [{Bos()}]{BoseQO}%
  \BibitemOpen
  \href@noop {} {\ }\bibinfo {note} {Whether superconducting coherence is
  destroyed thermally at $T>T_c$ or by vortex proliferation for $H>H_{c2}$, in
  the BEC limit the ``normal'' state proximate to the superconducting state is
  an uncondensed phase of ``preformed'' Cooper pairs, with vanishing density of
  electron quasi-particles.}\BibitemShut {Stop}%
\bibitem [{x_a()}]{x_and_EF}%
  \BibitemOpen
  \href@noop {} {\ }\bibinfo {note} {$E_F$ can only be directly related to the
  electron density in cases in which the band-structure is well approximated by
  a free electron dispersion, which is not the case in the cuprates. The values
  of $E_F$ that we quote are obtained directly from photoemission so the issue
  of whether the electron density is $x$ or $(1+x)$ need not be resolved for
  our purposes.}\BibitemShut {Stop}%
\end{thebibliography}
%

\end{document}